\documentclass[aps,prl,reprint,longbibliography,nofootinbib]{revtex4-1}

\usepackage{amsmath,amssymb,bm,graphicx}
\usepackage[colorlinks=true,citecolor=blue,linkcolor=blue,urlcolor=blue]{hyperref}

\newcommand{\be}{\begin{equation}}
\newcommand{\ee}{\end{equation}}

\begin{document}

\title{Chiral Doubling of Heavy-Light Hadrons and the New Beauty--Strange  $B_{s0}^*(5700)^0$ Candidate}

\author{Maciej A. Nowak}
\affiliation{Institute of Theoretical Physics and Mark Kac Center for Complex Systems Research,\\
Jagiellonian University, Kraków, Poland}

\author{Ismail Zahed}
\affiliation{Center for Nuclear Theory, Department of Physics and Astronomy,
Stony Brook University, Stony Brook, New York 11794, USA}

\date{\today}

%%%%%%%%%%%%%%%%%%%%%%%%%%%%%%%%%%%%%%%%%%%%%%%%%%%%%%%%%%%%%%%%%%%%%%%%%%%%%%
\begin{abstract}
%%%%%%%%%%%%%%%%%%%%%%%%%%%%%%%%%%%%%%%%%%%%%%%%%%%%%%%%%%%%%%%%%%%%%%%%%%%%%%

The recent observation by the LHCb Collaboration of a new
beauty--strange meson $B_{s0}^*(5700)^0$~\cite{LHCb:2026Bs0star}, under the conditional assignment $J^P=0^+$ provides an important new test of the chiral
organization of heavy-light hadrons. More than three decades ago it
was proposed that the coexistence of heavy-quark spin symmetry and
spontaneously broken chiral symmetry implies that every heavy-light
spin multiplet should possess an opposite-parity partner separated by
a nearly universal mass gap determined primarily by the dynamics of
the light degrees of freedom~\cite{Nowak:1993vc,Bardeen:1993ae}. This chiral doubling scenario was later
developed into a quantitative heavy-hadron effective theory and
extended to the complete charm and beauty spectra. In particular, the
strange beauty sector was predicted to exhibit a parity splitting
closely related to that of the strange charm sector, with only small
$1/m_Q$ corrections. We briefly review the symmetry origin of chiral
doubling, derive the universal parity splitting through the
heavy-hadron chiral effective theory, and discuss the new LHCb
observation in the context of the original theoretical predictions.
Taken together with the established charm spectrum, if the new
beauty-strange state is confirmed to have $J^P=0^+$, its mass is consistent with its interpretation as the chiral partner of the ground state $B_s^0$. Finally, we stress that the above confirmation of the spin and parity   imposes in the chiral doubling scenario an existence of yet unobserved, very narrow beauty-strange meson with assignment $1^+$ at 5747 $\pm$ 2 MeV.  This work is dedicated to the memory of our friend and
collaborator  Mannque Rho (1936--2026).

%%%%%%%%%%%%%%%%%%%%%%%%%%%%%%%%%%%%%%%%%%%%%%%%%%%%%%%%%%%%%%%%%%%%%%%%%%%%%%
\end{abstract}
%%%%%%%%%%%%%%%%%%%%%%%%%%%%%%%%%%%%%%%%%%%%%%%%%%%%%%%%%%%%%%%%%%%%%%%%%%%%%%

\maketitle

%\begin{center}
%\emph{This work is dedicated to the memory of our friend and
%colleague Mannque Rho (1936--2026).}
%\end{center}

\vspace{1em}

%%%%%%%%%%%%%%%%%%%%%%%%%%%%%%%%%%%%%%%%%%%%%%%%%%%%%%%%%%%%%%%%%%%%%%%%%%%%%%
\section{Introduction}
%%%%%%%%%%%%%%%%%%%%%%%%%%%%%%%%%%%%%%%%%%%%%%%%%%%%%%%%%%%%%%%%%%%%%%%%%%%%%%

Heavy-light hadrons provide a unique window on the nonperturbative
structure of Quantum Chromodynamics (QCD), where two approximate
symmetries coexist over widely separated energy scales. 
In the limit of an infinitely heavy quark, the heavy quark behaves as
a static color source and its spin and flavor decouple from the
long-distance dynamics of the light degrees of freedom. The physical
origin of this approximate symmetry was first recognized by
Shuryak~\cite{Shuryak:1980,Shuryak:1982}, and subsequently developed into the
heavy-quark symmetry of Isgur and Wise
\cite{Isgur:1989vq,Isgur:1989ed} and the systematic heavy-quark
effective theory (HQET) of Georgi and others
\cite{Georgi:1990um,Neubert:1993mb}.
%In the limit
%of a heavy quark,
%$m_Q\gg\Lambda_{\rm QCD}$, the heavy quark behaves as a static color
%source and the dynamics becomes independent of its spin and flavor,
%leading to heavy-quark symmetry and its effective field theory
%description (HQET)~\cite{Georgi:1990um,Isgur:1989vq,Isgur:1989ed,Neubert:1993mb}.
Conversely, the light quark remains governed by the spontaneous
breaking of chiral symmetry,
\begin{equation}
SU(3)_L\times SU(3)_R\rightarrow SU(3)_V,
\end{equation}
which generates the constituent quark mass and the Goldstone bosons
that dominate low-energy hadronic dynamics. Heavy-light hadrons
therefore provide an ideal laboratory in which the interplay between
heavy-quark symmetry and spontaneous chiral symmetry breaking can be
studied experimentally.

Heavy-quark symmetry organizes heavy-light mesons into nearly
degenerate spin multiplets labeled by the total angular momentum of
the light degrees of freedom,
\begin{equation}
{\bf j}_\ell={\bf L}+{\bf s}_\ell,
\end{equation}
with the lowest multiplet consisting of the familiar
$(0^-,1^-)$ states. The residual hyperfine splitting within each
doublet is suppressed by $1/m_Q$, while the dynamics of the light
quark is encoded systematically in heavy-hadron chiral effective
theory~\cite{Wise:1992hn,Burdman:1992gh,Yan:1992gz,Casalbuoni:1996pg}.

More than three decades ago it was realized that the simultaneous
realization of heavy-quark symmetry and spontaneous chiral symmetry
breaking within the  chiral models has a profound consequence for heavy-light spectroscopy.
The original works of
Refs.~\cite{Nowak:1993vc,Bardeen:1993ae,Nowak:2003ra,Bardeen:2003kt}
showed that every heavy-quark spin multiplet should possess an
opposite-parity partner, leading to a pattern of {\it chiral
doublers}. Since the corresponding mass splitting between the chiral doublers originates from the
light constituent dynamics rather than from the heavy quark itself,
the parity gap was predicted to depend only weakly on the heavy-quark
mass and therefore to remain approximately universal from charm to
beauty.

The discoveries of the narrow
$D_{s0}^*(2317)$ and $D_{s1}(2460)$ mesons by the
BaBar/CLEO Collaborations
provided the first striking experimental realization of this
picture~\cite{Aubert:2003fg,Besson:2003cp}. These observations
motivated a reformulation of the chiral doubling scenario within a
unified heavy-hadron chiral effective theory describing both the
negative- and positive-parity multiplets on equal footing
\cite{Nowak:2003ra,Bardeen:2003kt}. Besides deriving the generalized
Goldberger--Treiman relation for heavy-light hadrons, this framework
extended the analysis to the complete
$(D,D_s,B,B_s)$ spectrum and predicted that the parity splitting in
the strange beauty sector should closely follow that observed in the
strange charm sector, up to calculable corrections of order $1/m_Q$.

The recent observation by the LHCb Collaboration of a new
beauty--strange meson~\cite{LHCb:2026Bs0star} (under the conditional assignement $J^P=0^+$) provides the first opportunity to confront these
predictions directly in the beauty sector. Owing to the much larger
bottom-quark mass, heavy-quark symmetry is expected to hold even more
accurately than for charm, making the beauty-strange spectrum a
particularly clean testing ground for the chiral doubling mechanism.
Together with the established charm and strange-charm spectra, the new
state considerably strengthens the empirical evidence that spontaneous
chiral symmetry breaking remains the primary organizing principle of
heavy-light spectroscopy.

The purpose of this work is to place the new LHCb observation~\cite{LHCb:2026Bs0star} in the
broader context of the chiral doubling framework. Rather than
reviewing the extensive literature on heavy-light spectroscopy, we
briefly summarize the underlying symmetry arguments, derive the
universal parity splitting within heavy-hadron chiral effective
theory, and discuss the implications of the new beauty-strange state
for the spectroscopy of heavy-light hadrons. The emerging picture
illustrates the remarkable predictive power of combining heavy-quark
symmetry with spontaneous chiral symmetry breaking, an idea first
proposed more than thirty years ago and now realized across both the
charm and beauty sectors.

%%%%%%%%%%%%%%%%%%%%%%%%%%%%%%%%%%%%%%%%%%%%%%%%%%%%%%%%%%%%%%%%%%%%%%%%%%%%%%
\section{Heavy-Quark and Chiral Symmetries}
%%%%%%%%%%%%%%%%%%%%%%%%%%%%%%%%%%%%%%%%%%%%%%%%%%%%%%%%%%%%%%%%%%%%%%%%%%%%%%

The chiral doubling scenario follows directly from the simultaneous
realization of heavy-quark spin symmetry and spontaneously broken
chiral symmetry. Individually, neither symmetry predicts parity
doubling. Heavy-quark symmetry organizes hadrons into nearly
degenerate spin multiplets, while chiral symmetry governs the dynamics
of the light constituent quark. Together they imply the existence of
opposite-parity partners whose splitting is controlled by the
spontaneous breaking of chiral symmetry.

For a heavy quark of mass
$m_Q\gg\Lambda_{\rm QCD}$, the momentum may be decomposed as
\begin{equation}
p_Q^\mu=m_Qv^\mu+k^\mu,
\end{equation}
where $v^\mu$ is the conserved hadron four-velocity and
$k^\mu={\cal O}(\Lambda_{\rm QCD})$ is the residual momentum. Removing
the heavy mass dependence through
\begin{equation}
Q(x)=e^{-im_Qv\cdot x}h_v(x),
\end{equation}
the leading heavy-quark effective Lagrangian becomes
\begin{equation}
{\cal L}_{\rm HQET}
=
\bar h_v\,iv\!\cdot\!D\,h_v
+
{\cal O}(1/m_Q),
\end{equation}
which is independent of the heavy-quark spin. Consequently, the
spectrum is classified by the total angular momentum carried by the
light degrees of freedom,
\begin{equation}
{\bf j}_\ell
=
{\bf L}
+
{\bf s}_\ell,
\end{equation}
and each value of $j_\ell$ generates a heavy-quark spin doublet with
total spin $J=j_\ell\pm\frac12$. For $j_\ell=\frac12$, the ground-state
doublet is
\begin{equation}
(0^-,1^-),
\end{equation}
which becomes degenerate in the heavy-quark limit.

The light-quark sector possesses an approximate
$SU(3)_L\times SU(3)_R$ chiral symmetry, spontaneously broken to
$SU(3)_V$. Following the nonlinear realization of
Ref.~\cite{Nowak:1993vc}, one introduces
\begin{equation}
\Sigma=\xi_L^\dagger\xi_R ,
\end{equation}
with
\begin{equation}
\xi_L
\rightarrow
h(x)\,\xi_L\,g_L^\dagger,
\qquad
\xi_R
\rightarrow
h(x)\,\xi_R\,g_R^\dagger,
\end{equation}
where $g_{L,R}\in SU(3)_{L,R}$ are global chiral transformations and
$h(x)\in SU(3)_V$ is the compensating hidden-local transformation. In
the unitary gauge,
\begin{equation}
\xi_L^\dagger=\xi_R\equiv\xi,
\qquad
\xi^2
=
\exp\!\left(\frac{i\Pi}{f_\pi}\right).
\end{equation}

The negative-parity heavy-light multiplet is represented by the
superfield~\cite{Nowak:1993vc}
\begin{equation}
H
=
\frac{1+v\!\!\!/}{2}
\left(
P_\mu^*\gamma^\mu
-
P\gamma_5
\right),
\label{Hfield}
\end{equation}
while the positive-parity multiplet is described by
\begin{equation}
G
=
\frac{1+v\!\!\!/}{2}
\left(
P_{1\mu}\gamma^\mu\gamma_5
-
P_0
\right).
\label{Gfield}
\end{equation}
Under heavy-quark spin symmetry,
\begin{equation}
H\rightarrow SH,
\qquad
G\rightarrow SG,
\end{equation}
where $S\in SU(2)_Q$ satisfies
$[S,v\!\!\!/\,]=0$. Under the nonlinear realization of chiral
symmetry,
\begin{equation}
H\rightarrow Hh^\dagger(x),
\qquad
G\rightarrow Gh^\dagger(x),
\end{equation}
so that the heavy-spin and light-flavor transformations act
independently on the heavy and light constituents.

The chiral vector and axial currents are
\begin{align}
V_\mu
&=
\frac{i}{2}
\left(
\xi^\dagger\partial_\mu\xi
+
\xi\partial_\mu\xi^\dagger
\right),\\
A_\mu
&=
\frac{i}{2}
\left(
\xi^\dagger\partial_\mu\xi
-
\xi\partial_\mu\xi^\dagger
\right),
\end{align}
and the covariant derivative is
\begin{equation}
D_\mu H
=
\partial_\mu H
+
H V_\mu,
\qquad
D_\mu G
=
\partial_\mu G
+
G V_\mu.
\end{equation}
To leading order the effective Lagrangian is
\begin{align}
{\cal L}
=&
-i\,{\rm Tr}
\left(
\bar Hv\!\cdot\!DH
+
\bar Gv\!\cdot\!DG
\right)
\nonumber\\
&
+
g_H
{\rm Tr}
\left(
\bar HH\gamma_\mu\gamma_5A^\mu
\right)
+
g_G
{\rm Tr}
\left(
\bar GG\gamma_\mu\gamma_5A^\mu
\right)
\nonumber\\
&
+
g_{HG}
{\rm Tr}
\left[
\gamma_5
(\bar GH-\bar HG)
\gamma_\mu A^\mu
\right],
\label{lag}
\end{align}
where residual mass terms and corrections of order $1/m_Q$ have been
suppressed.

The coexistence of the $H$ and $G$ multiplets is the defining feature
of chiral doubling~\cite{Nowak:1993vc,Bardeen:1993ae}. Heavy-quark
symmetry guarantees the spin degeneracy within each multiplet,
whereas spontaneous chiral symmetry breaking determines the separation
between the two parity multiplets. Since this splitting originates in
the light-quark sector, it depends only weakly on the heavy-quark
mass, receiving corrections of order
\begin{equation}
{\cal O}\!\left(\frac{c_s}{m_Q}\right),
\end{equation}
where $c_s$ has dimensions of mass squared. This simple observation
explains why parity splittings are approximately similar, modulo
$SU(3)$ flavor breaking and strange-heavy mixing effects, throughout
the $D$, $D_s$, $B$, and $B_s$ spectra. The resulting chiral mass gap
is one of the central predictions of the chiral-doubling framework and
provides the basis for interpreting the newly observed beauty-strange
state as the positive-parity partner of the ground-state $B_s$
multiplet.

%%%%%%%%%%%%%%%%%%%%%%%%%%%%%%%%%%%%%%%%%%%%%%%%%%%%%%%%%%%%%%%%%%%%%%%%%%%%%%
\section{Chiral Doubling and the Universal Mass Gap}
%%%%%%%%%%%%%%%%%%%%%%%%%%%%%%%%%%%%%%%%%%%%%%%%%%%%%%%%%%%%%%%%%%%%%%%%%%%%%%

The physical heavy-meson multiplets $H$ and $G$ introduced in the
previous section transform nonlinearly under chiral symmetry.
It is convenient to combine these parity eigenstates into the
left- and right-handed fields
\begin{equation}
{\cal H}_L
=
\frac{1}{\sqrt2}
\left(
H+\gamma_5G
\right),
\qquad
{\cal H}_R
=
\frac{1}{\sqrt2}
\left(
H-\gamma_5G
\right),
\label{HLR}
\end{equation}
which are exchanged under parity and transform identically,
\begin{equation}
{\cal H}_L
\rightarrow
S{\cal H}_Lh^\dagger(x),
\qquad
{\cal H}_R
\rightarrow
S{\cal H}_Rh^\dagger(x).
\end{equation}
In terms of these fields, the most general leading-order effective
Lagrangian consistent with heavy-quark spin symmetry and chiral
symmetry is
\begin{align}
{\cal L}
=&
-i\,{\rm Tr}
\left(
\bar{\cal H}_Lv\!\cdot\!D{\cal H}_L
+
\bar{\cal H}_Rv\!\cdot\!D{\cal H}_R
\right)
\nonumber\\
&
-\Delta\,
{\rm Tr}
\left(
\bar{\cal H}_L\Sigma^\dagger{\cal H}_R
+
\bar{\cal H}_R\Sigma{\cal H}_L
\right)
\nonumber\\
&
+g\,
{\rm Tr}
\left(
\bar{\cal H}_L\gamma_\mu\gamma_5A^\mu{\cal H}_L
-
\bar{\cal H}_R\gamma_\mu\gamma_5A^\mu{\cal H}_R
\right),
\label{LRlag}
\end{align}
where the second term mixes the left- and right-handed sectors through
the chiral order parameter.

After spontaneous chiral symmetry breaking,
\begin{equation}
\langle\Sigma\rangle=\mathbf1,
\end{equation}
this mixing generates opposite mass shifts for the parity eigenstates,
\begin{equation}
M_H=\bar M-\Delta,
\qquad
M_G=\bar M+\Delta,
\end{equation}
leading to the universal parity splitting
\begin{equation}
\Delta M
\equiv
M_G-M_H
=
2\Delta.
\label{gap}
\end{equation}
Equation~(\ref{gap}) is the central result of the chiral-doubling
framework. Since the heavy quark acts only as a static color source,
the parameter $\Delta$ is determined primarily by spontaneous chiral
symmetry breaking in the light-quark sector. Consequently,
\begin{equation}
\Delta M
=
2\Delta
+
{\cal O}
\left(
\frac{c_s}{m_Q}
\right),
\label{1overm}
\end{equation}
so that the parity splitting depends only weakly on the heavy-quark
flavor.

The same result follows from the off-diagonal axial coupling between
the physical $H$ and $G$ multiplets. Expanding the axial current to
leading order,
\begin{equation}
A_\mu
=
-\frac{1}{2f_\pi}
\partial_\mu\Pi
+\cdots,
\end{equation}
the interaction between opposite-parity multiplets becomes
\begin{equation}
{\cal L}_{HG\pi}
=
\frac{g_{HG}}{2f_\pi}
{\rm Tr}
\left(
\gamma_\mu\gamma_5
(\bar GH-\bar HG)
\partial^\mu\Pi
\right).
\end{equation}
Integrating by parts and using the on-shell residual equations of
motion for the $H$ and $G$ fields yields the non-derivative pion
coupling
\begin{equation}
g_{\pi HG}
=
\frac{g_{HG}}{2f_\pi}
(M_G-M_H),
\label{GT}
\end{equation}
which is the generalized Goldberger--Treiman
relation~\cite{Nowak:2003ra}.
The quark Goldberger--Treiman relation reads
\begin{equation}
\Sigma_q
=
\frac{f_\pi\,g_{\pi qq}}{g_A^{(q)}}\,,
\end{equation}
  The phenomenological matching  $g_{\pi HG}=\frac{1}{2} g_{\pi qq}$, together with $g_{GH}=g_A^{(q)}$ leads to $M_G-M_H \sim \Sigma_q$. 
This provides a dynamical
explanation for the near universality of the mass gap across heavy
flavors. Comparison of $qq\pi$ and $\pi HG$ couplings requires a comment. The first one is generated by the local operator, whereas the second is a transition between two spatially extended bound states. The heavy quark acts as a static color source that reorganizes light quark cloud. Projecting the local $\pi qq$ vertex onto the lowest opposite parity heavy-light states produces a transition form factor smaller than unity, naturally of order one half. Qualitative support of such on overlap suppression is known in the literature~\cite{GoityRoberts}.

An immediate consequence is that the strange beauty and strange charm
sectors should exhibit nearly identical parity splittings,
\begin{equation}
M(B_s^{0^+})-M(B_s^{0^-})
\simeq
M(D_s^{0^+})-M(D_s^{0^-}),
\end{equation}
and similarly for the $(1^+,1^-)$ multiplets, with deviations
suppressed by inverse powers of the heavy-quark mass. This prediction,
formulated qualitatively in the original chiral doubling papers and
developed quantitatively within the heavy-hadron chiral effective
theory, provides the theoretical framework for interpreting the newly
observed beauty-strange state.

%%%%%%%%%%%%%%%%%%%%%%%%%%%%%%%%%%%%%%%%%%%%%%%%%%%%%%%%%%%%%%%%%%%%%%%%%%%%%%
\section{The New Beauty--Strange State}
%%%%%%%%%%%%%%%%%%%%%%%%%%%%%%%%%%%%%%%%%%%%%%%%%%%%%%%%%%%%%%%%%%%%%%%%%%%%%%

The recent observation by the LHCb Collaboration of a new
beauty--strange meson~\cite{LHCb:2026Bs0star} provides the first opportunity to test the chiral doubling scenario in the beauty-strange sector.  While the charm sector has provided
compelling evidence for parity doubling through the discovery of the
$D_{s0}^*(2317)$ and $D_{s1}(2460)$ states, the beauty sector probes
the same mechanism in a regime where heavy-quark symmetry is expected
to be even more accurate. The new state therefore constitutes the
first direct test of the chiral doubling picture in the strange beauty
sector.

Within heavy-quark symmetry, the ground-state beauty-strange mesons
form the familiar spin doublet
\begin{equation}
(0^-,1^-),
\end{equation}
while chiral symmetry predicts the existence of the corresponding
positive-parity doublet
\begin{equation}
(0^+,1^+).
\end{equation}
The parity splitting between these multiplets is generated by the
spontaneous breaking of chiral symmetry and is therefore expected to
depend only weakly on the heavy-quark mass.

To leading order,
\begin{eqnarray}
\Delta M_{B_s}&=&\Delta M_{\infty} + \frac{c_s}{m_b} \\
\Delta M_{D_s} &=&\Delta M_{\infty} + \frac{c_s}{m_c}
\label{universality}
\end{eqnarray}
where $\Delta M_{\infty}$ represents the chiral shift in the infinitely heavy quark limit. 
In our 2003 paper~\cite{Nowak:2003ra}, we simply approximated  $\Delta M_{B_s} \approx \Delta M_{\infty} $ quoting the value $\Delta M_{\infty}=323\,\rm  MeV$. 
The argument for the value  $\Delta M_{\infty}$ was as follows: we used the result from the instanton liquid model~\cite{Pobylitsa:1989uq,Musakhanov2001} for the value of constituent mass as a function of the quark current mass. The constituent mass is the sum of two terms: the current mass ($m_s$) and the dynamical mass, generated by the spontaneous breakdown of  chiral symmetry, $\Sigma_{\rm {dyn}}=\Sigma_q(0)(\sqrt{1+(m_s/d)^2}-m_s/d)$, where $d$ is a function of instanton parameters. For the standard instanton parameters (size is 1/3 fm, inter-instanton distance 1 fm),  we have  $d=198$ MeV and  about standard constituent mass for massless quark $\Sigma(0)=345 \, \rm MeV$.  Then, for  $m_s=150-155\, \rm MeV$,  one gets the chiral shift $\Sigma_q(m_s)=m_s+\Sigma_q(0)/2$=321.5+323.5 MeV, which expians the suggested  gap $323$ MeV in~\cite{Nowak:2003ra}.
Now, we can do better. 
Under the additional one-parameter assumption that the leading \(1/m_Q\)
correction is governed by the same coefficient \(c_s\)
for charm and beauty, eliminating \(c_s\) gives
\begin{equation}
\frac{\Delta M_{B_s} -\Delta M_{\infty}}{\Delta M_{D_s} -\Delta M_{\infty}}=\frac{m_c}{m_b}
\end{equation}
This equation can be tested as a consistency condition, e.g.  for the value of $\Delta M_{B_s}$:
\begin{equation}
\Delta M_{B_s} =\frac{m_c}{m_b}\Delta M_{D_s}+ \bigg(1-\frac{mc}{m_b}\bigg)\Delta M_{\infty}
\end{equation}
Using on the right-hand side the experimental value of $\Delta M_{D_s}$=349.4, the theoretical value   $\Delta M_{\infty}$= 323 MeV~\cite{Nowak:2003ra}, and the illustrative ratio $m_c/m_b=1/3$, we get the value 331.8 MeV, in remarkable agreement with the split between the ground state $B_s$ pseudoscalar and the just measured $B_{s0}(5700)^0$ by LHCb.

This prediction is considerably more restrictive than the existence of
an excited $B_s$ resonance alone. Heavy-light spectroscopy contains
many orbital and radial excitations, whereas the chiral doubling
framework predicts a specific parity structure tied directly to the
realization of chiral symmetry. The relevant comparison is therefore
between parity partners, also for different heavy flavors,  rather than between isolated masses.

%The physical origin of Eq.~(\ref{universality}) is transparent.
%Writing the heavy-light masses as
%\begin{equation}
%M_H
%=
%m_Q
%+
%\bar\Lambda
%-
%\Delta,
%\qquad
%M_G
%=
%m_Q
%+
%\bar\Lambda
%+
%\Delta,
%\end{equation}
%immediately yields
%\begin{equation}
%M_G-M_H
%=
%2\Delta,
%\end{equation}
%where the heavy-quark mass cancels identically at leading order.
%Finite-mass effects enter only through the higher-order HQET
%expansion,
%\begin{equation}
%M_G-M_H
%=
%2\Delta
%+
%\frac{c}{m_Q}
%+
%{\cal O}
%\left(
%\frac1{m_Q^2}
%\right),
%\end{equation}
%which explains why the parity splitting should remain approximately
%constant from charm to beauty.

%The generalized Goldberger--Treiman relation derived in the previous
%section provides an equivalent interpretation. Since
%\begin{equation}
%M_G-M_H
%=
%g_\pi f_\pi
%+
%{\cal O}\bigg(\frac 1{m_Q}\bigg),
%\end{equation}
%the universal parity gap reflects the same light-quark dynamics that
%governs the coupling of the Goldstone bosons to the heavy-light
%multiplets. The heavy quark merely acts as a static color source and
%plays little role in determining the magnitude of the splitting.

The present experimental situation therefore provides a particularly
stringent test of the chiral doubling hypothesis. The strange-charm
sector first established the existence of low-lying positive-parity
partners, while the newly observed beauty-strange state extends this
pattern to substantially larger heavy-quark masses. Taken together,
the two sectors suggest that the parity splitting is controlled
primarily by spontaneous chiral symmetry breaking rather than by the
heavy-quark mass itself.

The comparison between theory and experiment is summarized in
Table~\ref{tab:comparison}. The table highlights the essential feature
of the chiral doubling framework: the relevant observable is the
parity splitting within a given heavy-light family, not the absolute
mass of an individual resonance.
\begin{table*}[t]
\centering
\caption{Comparison of the strange charm and strange beauty parity
doublets. The ground-state masses are taken from the Particle Data
Group, while the positive-parity beauty-strange mass is from the recent
LHCb observation of the $B_{s0}^{*}(5700)^0$
\cite{LHCb:2026Bs0star}. The before-last column gives the chiral-doubling
expectation of Nowak, Rho, and Zahed \cite{Nowak:2003ra}. The result in quadratic brackets comes from interpolation done in this work. The last column gives results from lattice simulations (for charmed mesons~\cite{Bali}, for beauty mesons~\cite{Lang}). The quoted
uncertainty on $\Delta M(B_s)$ is obtained by adding the statistical,
systematic, and $B_s$-mass uncertainties in quadrature.}
\label{tab:comparison}
\setlength{\tabcolsep}{9pt}
\renewcommand{\arraystretch}{1.15}
\begin{tabular}{lcccccc}
\hline\hline
Sector & $J$ &
$M(J^-)$ &
$M(J^+)$ or construction  &
{exp.  gap}  &
${\rm NRZ}$  & lattice
\\
\hline
$D_s$ & $0$ &
$1968.34\pm0.07$ &
$2317.8\pm0.5$ &
$349.46\pm0.5$ &
$\sim 345$ & $371(4)^{+6}_{-0}$ 
\\
$D_s^*$ & $1$ &$2112.2 \pm0.4$ & $2459.5  \pm0.6 $  & $347.3 \pm 0.7$ & $\sim 345$ & $356(4)^{+1}_{-0} $  \\
$B_s$ & $0$ &
$5366.92\pm0.10$ &
$5698.9\pm1.5\pm0.6$ &
$331.98\pm1.62$ if $0^+$ &$323  [331.8]$  & $344.1 \pm 13 \pm 19$
\\
$B_s^*$ & $1$ &
$5415.4\pm 1.4$ &
$5747.4 \pm 2.1  $ &
$331.98\pm 1.62$ by construction &
$323[331.8]$  & $334.6 \pm 17 \pm 19$ \\
\hline\hline
\end{tabular}
\end{table*}

The significance of the new beauty-strange state extends beyond the
identification of a single hadron. Together with the established charm
spectrum, it supports the broader picture that spontaneous chiral
symmetry breaking continues to organize heavy-light spectroscopy over
a wide range of heavy-quark masses. The remarkable persistence of the
same parity structure from charm to beauty is precisely the hallmark
anticipated by the chiral doubling scenario.

\section{The missing doubler}

An eventual  experimental confirmation that $B_s^{0*}$ (5700) is indeed a scalar,  implies the existence, in the chiral doublets scenario,  of narrow  beauty-strange meson  with assignement $1^+$. In Figure~1 we present schematically the chiral doubling picture in the lowest plateau of the light spin, i.e. $j_l=1/2$.  Since both splits (i.e. due to the heavy spin effects and due to the spontaneous breakdown of the chiral symmetry (SB$\chi$S) have been measured and yield 48.5 MeV and 332 MeV (respectively), the mass of the missing right upper corner can be precisely estimated as  5747 MeV, within the accuracy of 2-3 MeV.  
Repeating the consistency check for  averaged $\bar{M}_{H}(Q)=\frac{M_{0^-}(Q)+3M_{1^-}(Q)}{4}$ 
 and  $\bar{M}_{G}(Q)=\frac{M_{0^+}(Q)+3M_{1^+}(Q)}{4}$  for $Q=c,b$, does not change the predicion very much.  The advantage is that the leading chromomagnetic terms cancel in the average, which reduces the $1/m_Q$ correction. This leads to the value of the mass split between positive parity doublet equal to $46.92\,\rm MeV$, comparing to the observed  ground doublet value $M(B_s^*)-M(B_s)=48.5 \, \rm MeV$, so the "distortion"  of the exact chiral rectangle~(Fig.~1) is negligible (within 2 MeV).  
 
Since this still highly hypothetical state  $1^+$  is 71 MeV below $B^*K$, the state has to be  very narrow, with isospin violating hadronic mode (decay to $B_s^*\pi^0$) and electromagnetic transitions  to $B_s \gamma$ and   $B_s^* \gamma$.   

Other chiral doublers are also expected in the non-strange beauty
sector. As in the corresponding non-strange charmed system, these
states are expected to be broad because they can decay strongly
through isospin-conserving channels. Chiral doubling is not restricted
to the lowest $j_\ell=\frac12$ multiplet but extends naturally to the
orbitally excited heavy-light states with
$j_\ell=\frac32$~\cite{Nowak:1993zz}. In this case the effective
theory predicts a substantially smaller parity splitting than for the
ground-state multiplet, reflecting the reduced influence of
spontaneous chiral symmetry breaking on orbitally excited states. The
2003 analysis estimated a chiral splitting of order
$170\,\mathrm{MeV}$ for the excited doublets, roughly a factor of two
smaller than for the lowest-lying heavy-light mesons.  Finally, chiral doubling has also consequences for the baryons (conventional and exotic), as already pointed out in~\cite{NowakRhoZahed1996}.

   %\graphics[width=0.8\linewidth]{Missingcorner.png}

\begin{figure}
        \centering
        \includegraphics[width=0.8\linewidth]{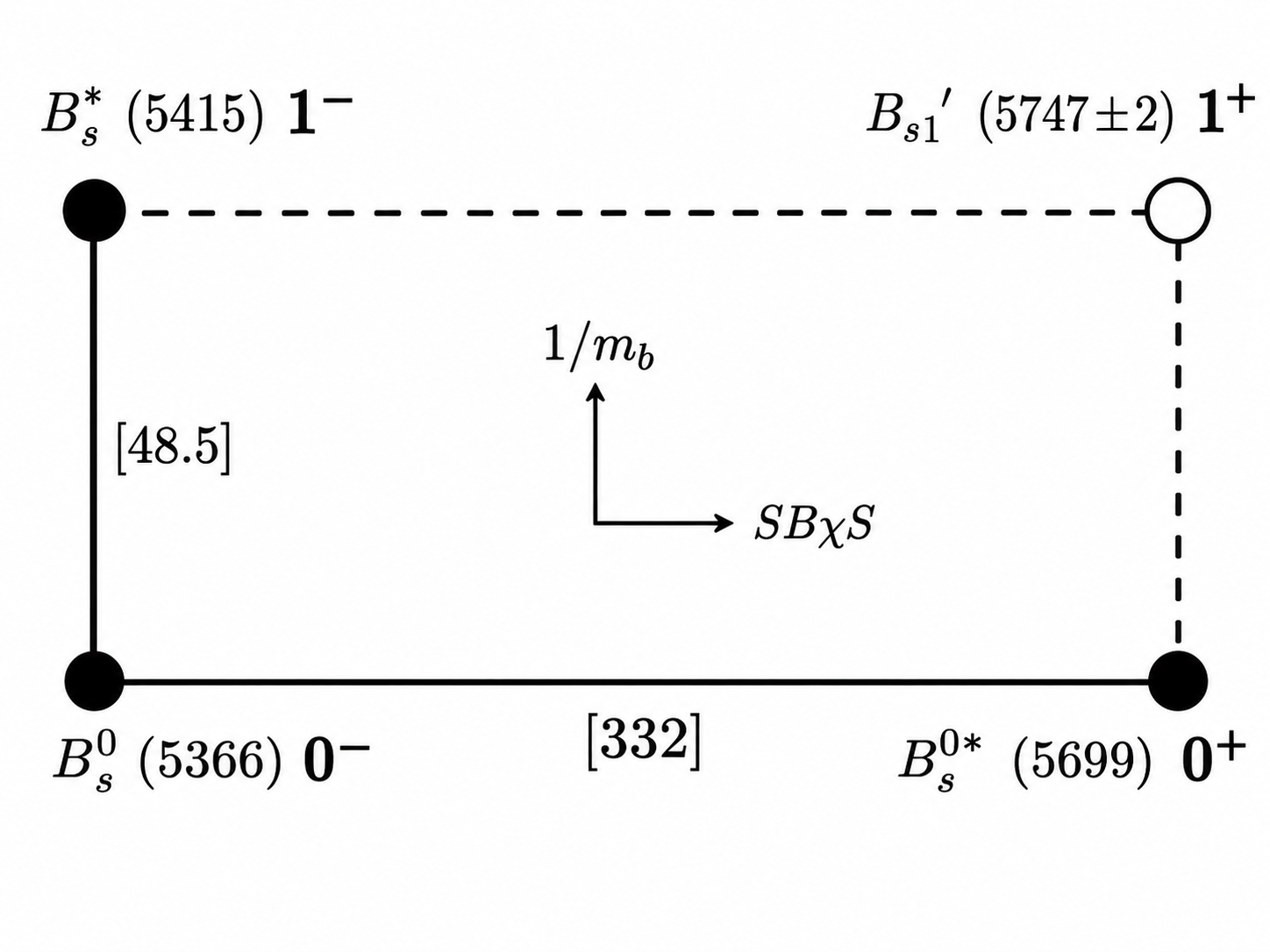}
        \caption{Lowest light spin plateau $j_l=1/2$ in the chiral doublers scenario for beauty-strange mesons. Full circles denote the observed particles, blank circle points at the missing, but expected one. Horizontal (vertical) lines represent chiral (heavy spin) shift, respectively. Note that both shifts $M(0^+)-M(0^-)\approx M(1^+)-M(1^-)$ and $M(1^+)-M(0^+)\approx M(1^-)-M(0^-)$ 
        lead, within experimental error of 2MeV, to the mass 5747 MeV. One should not confuse the upper right corner with the observed $B_{s1}(5830)^0$, which belongs to $j_l=3/2$ plateau. } 
        \label{fig:placeholder}
    \end{figure}

%%%%%%%%%%%%%%%%%%%%%%%%%%%%%%%%%%%%%%%%%%%%%%%%%%%%%%%%%%%%%%%%%%%%%%%%%%%%%%
\section{Discussion and Outlook}
%%%%%%%%%%%%%%%%%%%%%%%%%%%%%%%%%%%%%%%%%%%%%%%%%%%%%%%%%%%%%%%%%%%%%%%%%%%%%%

The observation of a new beauty--strange meson by the LHCb
Collaboration~\cite{LHCb:2026Bs0star} marks an important new stage in the experimental
exploration of heavy-light spectroscopy. Together with the established
positive-parity charmed and strange-charmed states, it can extend the
evidence for chiral doubling from the charm sector into the beauty
sector, where heavy-quark symmetry is expected to be realized with
even greater accuracy.

The central prediction of the chiral doubling framework~\cite{Nowak:1993vc,Bardeen:1993ae,Nowak:2003ra,Bardeen:2003kt} is not merely
the existence of additional heavy-light resonances, but the emergence
of a specific parity structure dictated by the simultaneous realization
of heavy-quark spin symmetry and spontaneous chiral symmetry breaking.
Heavy-quark symmetry organizes hadrons into nearly degenerate spin
multiplets, whereas spontaneous chiral symmetry breaking generates an
approximately universal mass gap between opposite-parity multiplets.
To leading order, this gap is independent on the heavy-quark mass and
is controlled instead by the dynamics of the light constituent quark.
The generalized Goldberger--Treiman relation provides the dynamical
connection between the parity splitting and the coupling of the
Goldstone bosons to heavy-light hadrons.

Historically, the idea that heavy-light hadrons should occur in
opposite-parity multiplets was proposed more than three decades ago on
symmetry grounds~\cite{Nowak:1993vc,Bardeen:1993ae,Nowak:2003ra,Bardeen:2003kt}. It was subsequently developed into a quantitative
heavy-hadron chiral effective theory encompassing the complete charm
and beauty spectra, where the near universality of the parity
splitting was identified as one of its principal phenomenological
predictions. The present observation provides (under the conditional assignment $J^P=0^+$) an important new
experimental support  for  this picture and considerably enlarges the
domain over which it can be tested.

Although the current experimental evidence strongly supports the
overall pattern predicted by chiral doubling, important questions
remain. The complete identification of the positive-parity beauty
multiplets, improved measurements of their masses, widths and quantum
numbers, and the spectroscopy of higher orbital and radial
excitations will further test the heavy-flavor independence of the
parity gap.
Since alternative scenarios for the $D_s$ multiplets were developed in the literature, either on the basis of molecular scenarios (so popular in the days of the abundance of exotics) or coupled channel dynamics, additional experimental evidence is needed, to characterize new state either as the  predominantly compact $q\bar{q}$  or the molecular (hadronic) one. From this perspective, the recent experimental result~\cite{BelleBelleII:2026Ds0Radiative}  for $D_{s0}^*(2317)^+ \rightarrow D_s^* \gamma$ is particularly encouraging, since the new state $B_s(5700)$ may have similar chiral content (modulo replacement of c quark by b), but very different electromagnetic properties, due to the different charges  of quarks b and c. Careful examination of all possible channels of decay can therefore  help to discriminate between  compact/molecular nature of the new state.    Equally important will be increasingly precise lattice QCD
calculations, which can determine the size of the expected
$1/m_Q$ corrections and establish the relation between the parity
splitting and the underlying pattern of spontaneous chiral symmetry
breaking directly from QCD.

Viewed in retrospect, the evolution of heavy-light spectroscopy has
been remarkable. What began as a symmetry argument based on the
coexistence of heavy-quark and chiral symmetries has evolved into a
coherent phenomenological framework spanning the charm and beauty
sectors. The new beauty--strange state reported by LHCb may serve as  an
important milestone in this development and seems to provide renewed evidence
that spontaneous chiral symmetry breaking continues to play an important  role in organizing the spectrum of hadrons containing a
single heavy quark.

%%%%%%%%%%%%%%%%%%%%%%%%%%%%%%%%%%%%%%%%%%%%%%%%%%%%%%%%%%%%%%%%%%%%%%%%%%%%%%
\section*{Acknowledgments}
%%%%%%%%%%%%%%%%%%%%%%%%%%%%%%%%%%%%%%%%%%%%%%%%%%%%%%%%%%%%%%%%%%%%%%%%%%%%%%
Our initial (1993) and later works (2003)  on chiral doublers were done in collaboration with Mannque Rho, a friend, mentor and collaborator. With a heavy heart we note his passing at the beginning of this year. His dedication and drive contributed importantly to the success of the chiral doublers idea.  
A search for the narrow $B_{s1}'$ state near $5747$ MeV therefore offers a direct continuation of the chiral doubling program developed with Mannque Rho and a particularly fitting experimental test of his physical insight. \\  \\   
This work is supported in part by the U.S. Department of Energy under Contract No.~DE-FG02-88ER40388.
 M.A.N. is  supported by the Priority Research Area DigiWorld under the
Strategic Programme Excellence Initiative at the Jagiellonian University

\bibliography{LHCB-REFS}

\end{document}